\documentclass[prl,twocolumn,showpacs,superscriptaddress]{revtex4}
\pdfoutput=1
\usepackage{amssymb}
\usepackage{graphicx,amsmath}

\usepackage{bm,color,mathrsfs,hyperref}

\hypersetup{
    colorlinks=true, 
    linkcolor=blue,  
    citecolor=blue,  
}

\newcommand{\bea}{\begin{eqnarray}}
\newcommand{\eea}{\end{eqnarray}}
\newcommand{\beq}{\begin{equation}}
\newcommand{\eeq}{\end{equation}}
\newcommand{\benu}{\begin{enumerate}}
\newcommand{\enu}{\end{enumerate}}

\newcommand{\si}{\sigma}

\newcommand{\bL}{{\bf L}}

\begin{document}
\title{
Effect of uniaxial pressure on the magneto-structural transitions
\\
of iron arsenide superconductors
}

\date{\today}

\author{A. Cano}
\affiliation{ European Synchrotron Radiation Facility, 6 rue Jules
Horowitz, BP 220, 38043 Grenoble, France}

\author{I. Paul}
\affiliation{ Institut N\'{e}el, CNRS/UJF, 25 avenue des Martyrs, BP
166, 38042 Grenoble, France}

\begin{abstract}
We study theoretically the variation of the magnetic and the structural transition temperatures of the iron
arsenide superconductors with external uniaxial stress. We show that the
increase of the transition temperatures reported in recent experiments is compatible with a simple magneto-elastic model
in which the physical origin of this variation is linked to the fact that antiferromagnetic bonds are longer than the
ferromagnetic ones in the magnetic phase. We make predictions which can be verified in order to test the relevance of this model.
\end{abstract}

\maketitle

\emph{Introduction}.
At low doping, iron arsenide superconductors
exhibit magneto-structural transitions from paramagnetic metals with tetragonal crystalline symmetry to low temperature
antiferromagnetic metals with $(\pi, 0)$ magnetic order  and orthorhombic crystal structure
(using the notation of 1Fe/cell Brillouin zone) \cite{reviews}.
These transitions are either concomitant or close to one another,
thereby suggesting the presence of non-trivial magneto-elastic coupling \cite{prb1,prb2}.
The structural transition consists of a $C_4$-symmetry breaking that, from a purely electronic point of view, can be seen as a
nematic transition \cite{hu11}.
Consistent with this interpretation, various experiments have reported in-plane anisotropy of the electronic properties
between the $x$- and the $y$- axes in the orthorhombic phase \cite{exp-ani}.
Since the study of such anisotropy is best performed using detwinned samples,
experimental methods have been developed to perform the detwinning mechanically using in-plane uniaxial pressure \cite{mech-stress}.
However, an important prerequisite for interpreting these data is the understanding of how the magneto-structural transitions
are themselves affected by the uniaxial pressure. Very recently experiments
on BaFe$_2$As$_2$, one of the most extensively studied systems for electron anisotropy,
have shown that the both the structural and the magnetic transition temperatures,  $T_S$ and $T_N$ respectively,
are remarkably sensitive to uniaxial pressure \cite{dhital11,blomberg}.
In fact, it has been reported that a modest compressive stress of 0.7 MPa along the shorter
$y$-axis results in an increase of $T_S$ by about 10 K, and that of $T_N$ by few Kelvins \cite{dhital11}.
Strikingly enough, tensile stress produces a similar increase of the magneto-structural transition temperatures \cite{blomberg}. This rules out a trivial magnetostriction effect because, in that case, compressive and tensile stresses are expected to give opposite (increase vs. decrease) results.

The purpose of this paper is to point out that these
changes in $T_S$ and $T_N$ can be understood quite readily as a consequence of magneto-elastic coupling.
Specifically, of the coupling that in the magnetic phase ensures that the antiferromagnetic
bonds are longer compared to their ferromagnetic counterparts, a feature which is universal
to all known iron based superconductors \cite{reviews}.
The magnitudes of the above changes are expected to increase if the system is near a second order
orthorhombic instability, which is the case of BaFe$_2$As$_2$.
As a corollary to these reasonings, we predict that applying tensile/compressive stress along
the shorter/longer Fe-Fe bonds in the orthorhombic phase will, in fact, reduce the magneto-structural transition temperatures.

\emph{Theory}.
We write the free energy of the system as
\[
F = F_M + F_E + F_{ME} - \si_{xx} u_{xx} - \si_{yy} u_{yy},
\]
where $\si_{ii}$ denotes external uniaxial stress, with the convention that $\si < 0$ for compressive stress
and $\si > 0$ for tensile stress,
and $u_{ij}$ are the components of the
in-plane strain tensor with $(i,j) = (x, y)$.
The magnetic part of the free energy is given by
\[
F_M = A \left( \bL_+^2 + \bL_-^2 \right)/2 + \cdots,
\]
where $\bL_+$ and $\bL_-$ are magnetic order parameters corresponding
to $(\pi, 0)$ and $(0, \pi)$ orders respectively, and $A = \alpha (T - T_N^0)$
where $T_N^0$ is the nominal N\'{e}el transition temperature. The elastic
part is described by
\bea
F_E &=& \frac{K}{2} \left( u_{xx} + u_{yy} \right)^2 + \frac{C_0}{2} \left( u_{xx} - u_{yy} \right)^2
\nonumber \\
&+& \frac{B}{4} \left( u_{xx} - u_{yy} \right)^4 + \cdots, \nonumber
\eea
where
$K$ is the bulk modulus,
$C_0$ is the temperature ($T$) dependent
orthorhombic elastic constant, and $B$ is a temperature independent constant.
The important magneto-elastic part is
given by \beq \label{eq:FME}
F_{ME} = -g \left( u_{xx} - u_{yy} \right) \left(\bL_+^2 - \bL_-^2 \right),
\eeq
with the coupling constant $g > 0$ such that in the magnetic phase the antiferromagnetic bonds are longer
than the ferromagnetic bonds
(see below).
The ellipses in the above equations denote terms that are not relevant for the current discussion.
Since the systems of interest are near a second-order structural transition where $C_0 \to 0$,
magnetostriction effects are sub-dominant and can be ignored.

The coupling \eqref{eq:FME} has been shown to play a key role in
establishing a universal phase diagram
of these systems \cite{prb1,prb2}.
In particular, the presence of tricritical points at which the magnetic transition changes character from first to
second order were predicted for this phase diagram in \cite{prb1}, and it has later
been confirmed experimentally in \cite{kim11}. However, in the following
we consider the case where the intrinsic instabilities (defined as the temperatures where the
coefficients $A$ and $C_0$ are zero) are sufficiently apart in $T$
such that the magnetic transition remains second order
and the elastic response to external stress is linear at $T_N$. For the sake of concreteness
we treat the case where $T_S > T_N$ (which is relevant for BaFe$_2$As$_2$). In the absence of
external stress the magnetic ordering takes place at
\beq
T_N (\si = 0) \equiv T_N (0) = T_N^0 + 2g \delta /\alpha, \nonumber
\eeq
where
$\delta \equiv ( |C_0| /B)^{1/2}$ is the spontaneous orthorhombic distortion that appears for $T< T_S$
due to the lattice instability ($C_0<0$).
Note that $g>0$ ensures that in the magnetic phase the antiferronagnetic bonds are longer
than the ferromagnetic ones (mathematically, the $u_{xx} > u_{yy}$ and the $u_{xx} < u_{yy}$ solutions
are coupled to the instabilities of $\bL_+$ and $\bL_-$ respectively). In the following we describe
two different experimental paths, namely stress applied above and below $T_S$ (in the presence of finite stress
$T_S$ defines a cross-over where $C_0$ vanishes),
and we assume that below $T_S$ the system is in a single domain state with $u_{xx} > u_{yy} $.

(i) \emph{Uniaxial stress applied in the paramagnetic tetragonal phase.} We consider
uniaxial stress $\si$  along one of the two equivalent directions, say $y$- for concreteness
as in \cite{dhital11}.
The strains induced by $\si$  are obtained by minimizing the total free energy ${\partial F \over \partial u_{ij}} =0$.
Since we are assuming that the system first feels the underlying structural instability and then undergoes the magnetic transition ($T_S > T_N$), it is convenient to carry out this minimization by expressing the total strain as $u_{ij} \to u_{ij}^0 + u_{ij} $, where $u_{ij}^0$ represent the strains that would appear spontaneously due to the structural transition at $T_S$ ($u_{xx}^0 - u_{yy}^0  = \pm \delta$).
In this way, the change in the effective orthorhombic elastic constant $C_0 \to \widetilde C_0 = 2|C_0|$ that takes place below $T_S$ is captured via $u_{ij}^0$.
Sufficiently below $T_S$
the elastic response is linear and the strains induced by $\sigma $
read
$u_{yy} = - (K + \widetilde C_0)/(K - \widetilde C_0) u_{xx} = \si (K + \widetilde C_0)/(4K \widetilde C_0)$.
Substituting these expressions back in the free energy we find that the effective magnetic energy is modified to
$F_M^{\prime} = F_M + g \si (\bL_+^2 - \bL_-^2)/(2\widetilde C_0)$.
The N\'{e}el temperature is
determined by the leading instability between $\bL_+$ and $\bL_-$
associated to $(\pi, 0)$ and $(0,\pi)$ orders respectively. In the case of compressive stress the $(\pi, 0)$ magnetic state is promoted, while with tensile stress
it is the $(0, \pi)$.
But for both compressive as well as for tensile stress $T_N$ increases
as
\beq
\label{eq:TN1}
T_N(\sigma) = T_N(0) + \frac{g |\si|}{\widetilde C_0 \alpha}.
\eeq
Note that this conclusion is also
valid
for systems where the intrinsic N\'{e}el
transition appears before the structural one (i.e., $T_N^0 > T_S$), with $\widetilde C_0$ replaced by $C_0$ in Eq.~\eqref{eq:TN1}.

(ii) \emph{Uniaxial stress applied in the paramagnetic orthorhombic phase.}
In the single-domain state with $u_{xx} > u_{yy}$, which promotes the $(\pi, 0)$ magnetic order,
the uniaxial stress can be applied either along the long or the short directions ($x$ and $y$ respectively).
The induced strains can be obtained by minimizing the total free energy as before. After substuting the resulting expressions back in the free energy we find that
stress applied along the short $y$ direction shifts the N\'{e}el temperature to
\beq
\label{eq:TN2}
T_N (\si) = T_N (0) - \frac{g \si}{\widetilde{C}_0 \alpha}.
\eeq
Thus, while $T_N$ increases in the case of compressive stress,
it actually decreases
in the case of tensile stress unlike in the previous case (i).
Obviously, the conclusion is reversed if the stress is applied along the long $x$-direction.

\begin{figure}[tb]
\includegraphics[width=.475\textwidth]{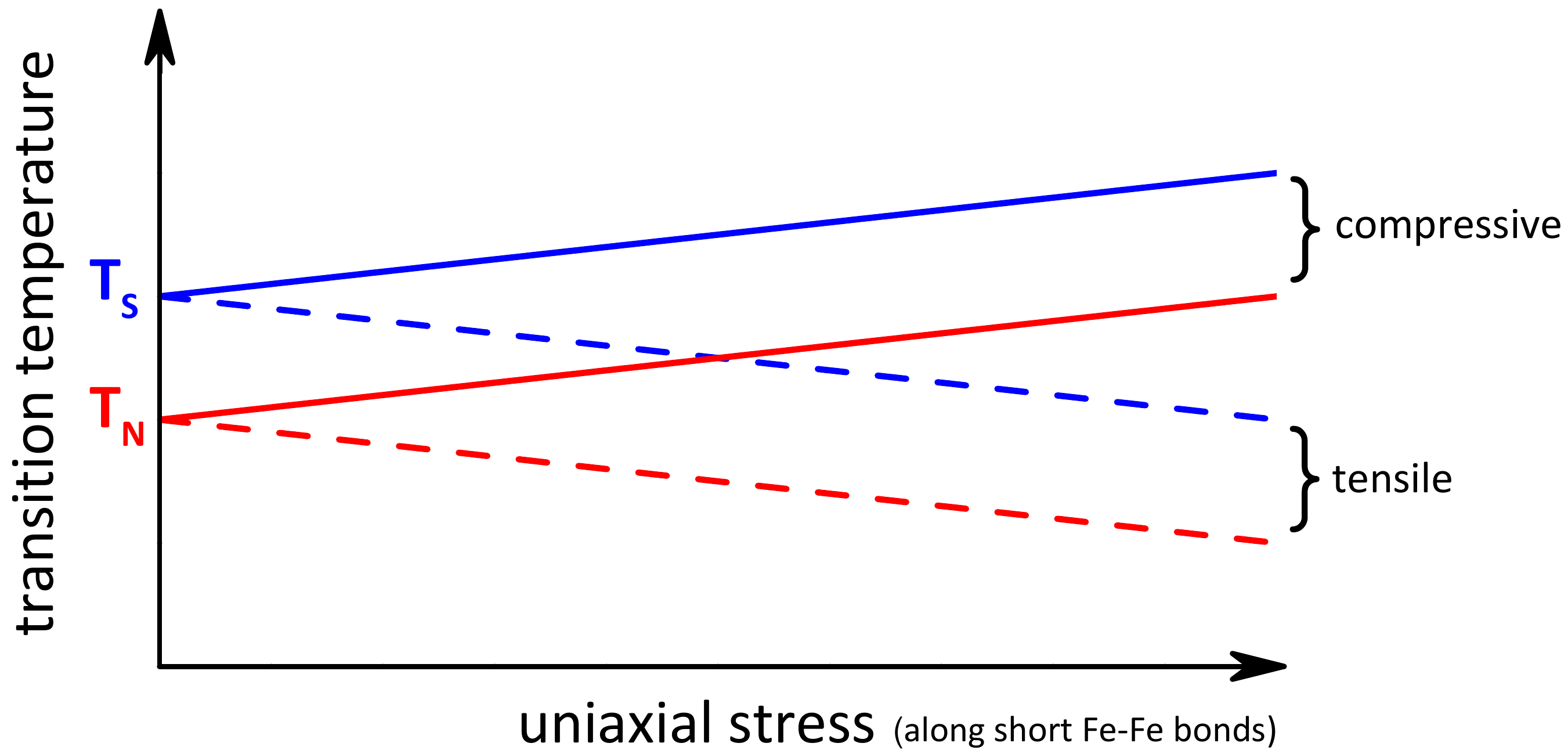}
\caption{
(Colour online). Schematic variation of magnetic and structural transition temperatures,
$T_N$ (red) and $T_S$ (blue) respectively, with uniaxial stress in detwinned samples.
According to Eq. \eqref{eq:TN1} $T_N$ increases, and $T_S$ tracks it (solid lines), for both
compressive and tensile stress applied in the tetragonal phase
as seen experimentally \cite{dhital11,blomberg}.
Eq. \eqref{eq:TN2} predicts that $T_N$ decreases, and $T_S$ tracks it (dash lines), for
tensile/compressive stress applied along the short/long Fe-Fe bonds in the orthorhombic
phase.}
\label{F1}
\end{figure}

\begin{figure}[tb]
\includegraphics[width=.4\textwidth]{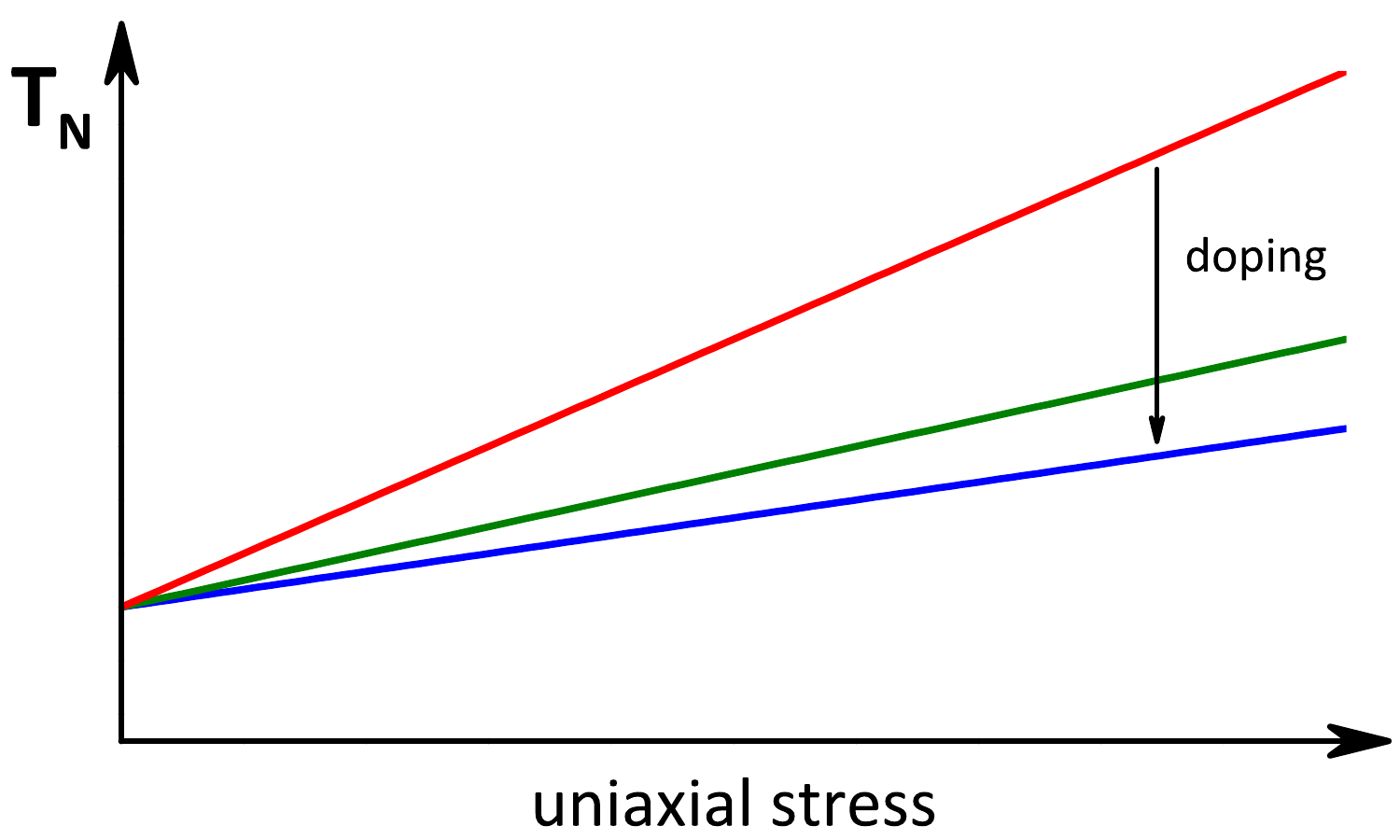}
\caption{
(Colour Online). Schematic variation of the magnetic transition temperature $T_N$ with uniaxial
stress $\sigma$ for different dopings. The slope $dT_N/d\sigma$, which is inversely
proportional to $|T_S - T_N|$ in the current theory, is predicted to decrease with doping.
}
\label{F2}
\end{figure}

\emph{Discussion}.
Eqs.~\eqref{eq:TN1} and \eqref{eq:TN2} are the main results of this paper, which are illustrated in Fig. \ref{F1}.
Note that these results are a direct consequence of
the magnetoelastic coupling  described by \eqref{eq:FME} and symmetry considerations.
As such, they do not depend upon whether the magnetic excitations are more itinerant-like
or more localized-like. Furthermore, it can be argued that these conclusions remain unchanged
even if the $C_4$-symmetry breaking is due to electron spin nematicity or due to orbital ordering
(in which case orthorhombicity is a secondary order parameter).
The linear
increase of $T_N$ with
both compressive and tensile external stress described by Eq. \eqref{eq:TN1}
has been observed experimentally
for small $\si$ \cite{dhital11,blomberg}.
Note that the sign of $g$ is crucial for understanding these experimental findings.
The decrease of $T_N$ with compressive/tensile uniaxial strain
applied along the long/short Fe-Fe bonds described by \eqref{eq:TN2},
on the contrary, is a prediction to be confirmed in future experiments. In fact, in the case where the
system is not completely detwinned, the applied
stress reduces the N\'{e}el temperature of the ``wrong'' domains compared to that of the
``correct'' ones, and this results in the smearing of the experimentally determined
$T_N$. Furthermore, since the orthorhombic elastic constant $C_0$ appears in the denominator in
both Eqs.~\eqref{eq:TN1} and \eqref{eq:TN2}, closer the
system is to a second order orthorhombic instability
as in BaFe$_2$As$_2$ and EuFe$_2$As$_2$ \cite{ying}, larger is the variation of $T_N$ with $\si$.
On the contrary, we expect the
magneto-structural transitions to be relatively insensitive to $\si$ for CaFe$_2$As$_2$ and SrFe$_2$As$_2$
for which the structural instability is first order. In fact, if we assume that
at the N\'{e}el transition $C_0$ varies linearly with temperature as $C_0 \propto T-T_S$, then we get that the
slope $dT_N/d\si$ varies inversely with $(T_S - T_N)$. This conclusion can in principle be tested by comparing the
results for samples at various doping as illustrated in Fig. \ref{F2}.

Finally, we
focus
on the variation of $T_S$ with external uniaxial
stress. It has been pointed out that the orthorhombic transition can
be interpreted as a result of structural softening due to magnetic
fluctuations that develop near a $(\pi, 0)$ magnetic order
\cite{prl,fernandes}. This interpretation provides a simple
explanation why $T_S$ and $T_N$ track each other in the
temperature-doping phase diagram of all the iron arsenides.
More concretely, if we assume that the structural transition occurs in a regime
where the magnetic fluctuations are controlled by the magnetic quantum
critical point, we get that the correction to the orthorhombic elastic constant
is~\cite{prl}
$ \delta C_0 = - \Lambda \ln[T_0/(T-T_N)]$, where $\Lambda$ and $T_0$ are constants
that depend on microscopic details.
The condition $C_0 + \delta C_0 =0$ determines the
structural transition point, from which we obtain
\beq
\label{eq:TS} T_S = T_N + T_0 e^{-C_0/\Lambda}. \eeq Thus, we expect
$T_S$ to increase (or decrease) with $\si$ in order to keep track of
the increase (or decrease) of $T_N$. Note that this conclusion is
qualitatively valid even if the magnetic fluctuations are controlled
by the finite temperature magnetic critical point.

\emph{Conclusions.}
We have studied the variation of the magneto-structural transition temperatures of the iron arsenide systems with uniaxial stress.
We have shown that the physical origin of this variation is the magneto-elastic coupling which makes the antiferromagnetic
bonds longer than the ferromagnetic ones in the magnetic phase.
The conclusions of the theory are compatible with the current experimental findings.
The magnitude of this variation
is relatively large because of the proximity of the orthorhombic instabilty, which is the case of BaFe$_2$As$_2$ (and presumably also of EuFe$_2$As$_2$).
In order to test this interpretation we predict that (i) the transition temperatures will decrease if,
after preparing the sample in a single-domain state,
tensile/compressive stress
is applied along the shorter/longer Fe-Fe bonds in the orthorhombic phase,
(ii) and that the slope $dT_N/d\si$ is inversely proportional to $|T_S -T_N|$ when the two transitions are
close enough (but not too close to make the magnetic transition first order).
Note that our results are independent of microscopic details concerning the magnetic properties
of the iron arsenides.

\begin{acknowledgments}
 We thank Y. Gallais, G. Garbarino and E. Kats for fruitful discussions.
\end{acknowledgments}

\emph{Note added}. Hu {\it et al.} \cite{hu12} have also studied pressure effects in iron-pnictides from a
microscopic spin model and a pure electron-nematic point of view. Their results are compatible with Eq. \eqref{eq:TN1}.

\end{document}